\documentclass[conference]{IEEEtran}
\IEEEoverridecommandlockouts
\usepackage[comma,sort&compress,numbers]{natbib}

\usepackage{amsmath,amssymb,amsfonts}
\usepackage{algorithmic,soul}
\usepackage{graphicx}
\usepackage{textcomp}
\usepackage{xcolor}
\usepackage{flushend}
\usepackage[
draft,
bookmarks=false,
pdfborder={0 0 0},
hidelinks
]{hyperref}
\usepackage{placeins}

\usepackage[font=small]{caption}

\usepackage{booktabs}
\usepackage{gensymb}
	\usepackage{multirow}
	\usepackage{sidecap}
    \usepackage{enumitem}

\usepackage[T1]{fontenc}
\usepackage{tgheros}

\def\BibTeX{{\rm B\kern-.05em{\sc i\kern-.025em b}\kern-.08em
    T\kern-.1667em\lower.7ex\hbox{E}\kern-.125emX}}
\begin{document}

\title{A Mamba-based Perceptual Loss Function for Learning-based UGC Transcoding
\thanks{The authors thank the funding from Tencent (US), University of Bristol, and the UKRI MyWorld Strength in Places Programme (SIPF00006/1).}
}

\author{\IEEEauthorblockN{Zihao Qi\textsuperscript{\dag}, Chen Feng\textsuperscript{\dag}, Fan Zhang\textsuperscript{\dag}, Xiaozhong Xu\textsuperscript{\S}, Shan Liu\textsuperscript{\S} and David Bull\textsuperscript{\dag}}
\IEEEauthorblockA{  \textsuperscript{\dag}\textit{Visual Information Laboratory, University of Bristol, Bristol, UK, BS1 5DD} \\
  \{zihao.qi, chen.feng, fan.zhang, dave.bull\}@bristol.ac.uk  \\
  \textsuperscript{\S}\textit{Tencent Media Lab, Palo Alto, CA 94306, USA} \\
  \{xiaozhongxu, shanl\}@tencent.com}
\vspace{-30pt}
}

% \author{
% \IEEEauthorblockN{Author(s) Name(s)}
% \IEEEauthorblockA{\textit{Author Affiliation(s)}}
% }

\maketitle

\begin{abstract}
In user-generated content (UGC) transcoding, source videos typically suffer various degradations due to prior compression, editing, or suboptimal capture conditions. Consequently, existing video compression paradigms that solely optimize for fidelity relative to the reference become suboptimal, as they force the codec to replicate the inherent artifacts of the non-pristine source. To address this, we propose a novel perceptually inspired loss function for learning-based UGC video transcoding that redefines the role of the reference video, shifting it from a ground-truth pixel anchor to an informative contextual guide. Specifically, we train a lightweight neural quality model based on a Selective Structured State-Space Model (Mamba) optimized using a weakly-supervised Siamese ranking strategy. The proposed model is then integrated into the rate-distortion optimization (RDO) process of two neural video codecs (DCVC and HiNeRV) as a loss function, aiming to generate reconstructed content with improved perceptual quality. Our experiments demonstrate that this framework achieves substantial coding gains over both autoencoder and implicit neural representation-based baselines, with 8.46\% and 12.89\% BD-rate savings, respectively.
\end{abstract}


\section{Introduction} 
\label{sec:intro}

\begin{figure}[!t]
  \centering
  \includegraphics[width=\linewidth]{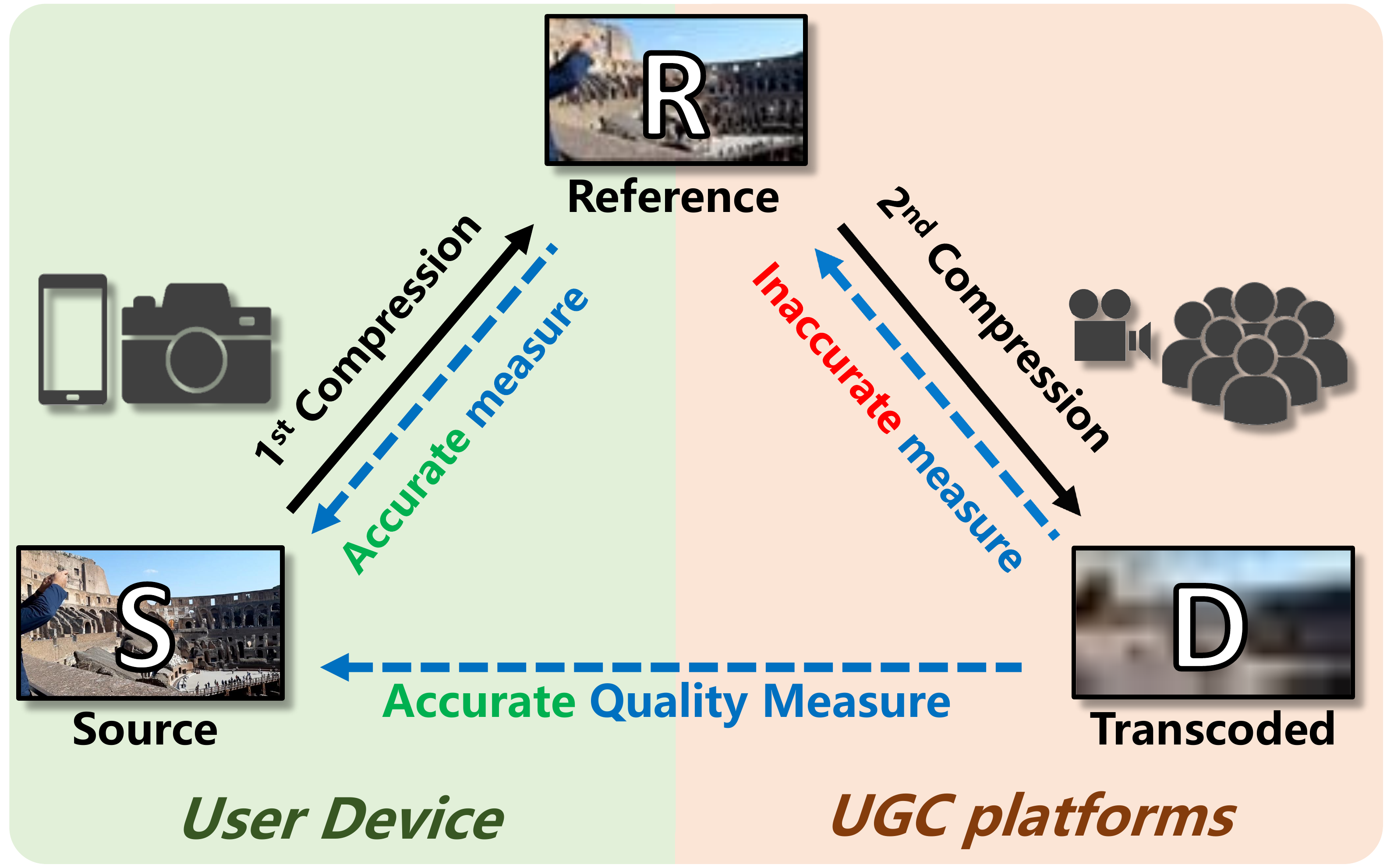}
  \caption{Illustration of the UGC video delivery pipeline. The $\mathbf{S}$ource content captured by a user is directly compressed on the user device for storage, as non-pristine $\mathbf{R}$eferences. The latter is then uploaded onto UGC streaming platforms and further compressed into $\mathbf{D}$istorted videos before being transmitted to the viewer.}
\label{fig:transcoding}\vspace{-10pt}
\end{figure}

The explosive growth of user-generated content (UGC) in recent years has fundamentally reshaped Internet video traffic. Industry reports~\cite{UGCPlatformsMarket2024} indicate that more than 500 hours of video are uploaded worldwide every minute, serving a user base that exceeds 1.06 billion. Unlike professionally-generated content (PGC), which originates from high-end capture devices with pristine master copies, UGC is typically captured by amateur users using mobile devices and undergoes lossy processing before it reaches a video sharing platform. As illustrated in \autoref{fig:transcoding}, a source video ($\mathbf{S}$) will typically be captured under suboptimal conditions and then compressed immediately on a mobile device to create a non-pristine reference ($\mathbf{R}$). This reference is then uploaded to and transcoded (to produce a distorted video $\mathbf{D}$) by UGC service providers, ready for streaming. It should be noted that the UGC transcoding process fundamentally differs from conventional coding as, in the former case, the input content ($\mathbf{R}$) is inherently degraded by prior processing.

Most existing compression frameworks (both conventional and learning-based) are fidelity-oriented, following a rate-distortion optimization (RDO) strategy with the primary objective of minimizing the difference between the reconstructed video and the input reference \cite{bull2021intelligent}. While this assumption is valid in cases where reference videos are pristine, it introduces a critical problem in UGC transcoding, because the reference video $\mathbf{R}$ often contains visible compression artifacts, noise, and/or editing flaws. The RDO process will force the codec to replicate these distortions, therefore resulting in low-quality reconstructions \cite{qi2024bvi}. 

To address this issue, based on recent advances in quality assessment for UGC transcoding \cite{qi2024full}, we propose a paradigm shift in the optimization objective for UGC transcoding. Here, we i) use the non-pristine reference as semantic context rather than as a pixel-level ground truth, and ii) introduce a perceptually-aware loss tailored for UGC transcoding. This new perceptual transcoding loss (PT-Loss) is based on a lightweight Selective Structured State-Space (Mamba) network, which has been trained via a weakly-supervised Siamese ranking strategy to predict perceptual quality degradation compared to pristine source content based only on non-pristine references and distorted video sequences. By integrating this loss into the transcoding process, a video codec can be guided to achieve improved coding gain over fidelity-based baselines. The main contributions of this paper are summarized below:

\begin{itemize}[leftmargin=*]
    \item \textbf{Perceptual loss for transcoding}: We propose a novel Mamba-based perceptual loss that treats the non-pristine reference as informative guidance rather than as a strict anchor. This allows the codec to deviate from reference fidelity to perceptual enhancement.  

    \item \textbf{Model-agnostic generalization}: We demonstrate the versatility of our loss by integrating it into two different neural video coding frameworks: an autoencoder-based codec (DCVC~\cite{li2021deep}) and an implicit neural representation (INR) based codec (HiNeRV~\cite{kwan2023hinerv}).

    \item \textbf{Significant coding gains}: Experimental results on a UGC transcoding dataset, BVI-UGC \cite{qi2024bvi}, show that both codecs with the proposed transcoding loss achieve consistent BD-rate gains (up to 12.89\% BD-rate) over the original baselines. Qualitative results also confirm the performance improvement from the perceptual perspective.
\end{itemize}

\section{Related Works}
\label{sec:related}

\subsection{Video Compression}

Conventional video compression remains widely used across most applications, with video coding standards evolving over the past few decades~\cite{h264_overview,h265_overview,h266_overview,chen2020overview}. Recently, however, learning-based coding tools and end-to-end neural video codecs have demonstrated their potential to compete with or even outperform conventional video codecs. While early approaches such as DVC~\cite{lu2019dvc} replace individual modules in the traditional compression pipeline with neural networks, more recent approaches extend this idea by adopting more advanced architectures, including deep context modeling~\cite{li2021deep}, hierarchical latent representations~\cite{lu2019dvc}, and implicit neural representations~\cite{chen2021nerv,kwan2023hinerv,gao2025givic}. It is noted that, for both conventional and learning-based methods, the underlying optimization principle remains the same -- the codec optimizes rate–distortion performance based on fidelity-driven quality metrics, such as PSNR or SSIM, which are computed between the reconstructed and reference videos. While this optimization strategy has proven effective for high-quality reference videos, it has been widely reported to pose a fundamental limitation for UGC video transcoding~\cite{qi2024bvi}. In such a scenario, the reference video itself already contains various visible artifacts, causing fidelity-driven optimization to preserve and even reinforce undesired distortions that exist in the reference content. 

\subsection{Video Quality Assessment}

Objective video quality assessment (VQA) methods are invariably used to guide the evaluation and optimization of video compression. Existing metrics can be divided into two major categories according to the availability of reference content: full-reference (FR) and no-reference (NR)\footnote{Although reduced reference (RR) VQA methods do exist in the literature, they are not commonly deployed in practical applications of video coding.}.

No-reference (NR) VQA metrics predict video quality without access to the reference. While effective for general quality estimation, it is difficult for NR models to measure fine-grained degradations relative to a source video~\cite{qi2024bvi}. Full-reference (FR) VQA, on the other hand, explicitly compares a video to its reference and has long been used to guide rate–distortion optimization in compression. Simple metrics like PSNR and SSIM~\cite{wang2004image,wang2003multiscale} provide straightforward fidelity measurements, while perceptually inspired metrics, such as VMAF~\cite{zhang2021enhancing}, incorporate human visual system characteristics to better align with subjective judgments. Recent deep learning-based FR models, including DeepVQA~\cite{kim2018deep}, LPIPS~\cite{zhang2018unreasonable}, C3DVQA~\cite{xu2020c3dvqa}, and Compressed UGC VQA~\cite{li2021full}, as well as unsupervised or weakly-supervised approaches like RankDVQA~\cite{feng2024rankdvqa}, have shown improved performance and generalization across datasets. However, existing FR methods have inherent limitations: they are optimized solely to approximate the reference, so the compression process guided by these metrics tends to reproduce the distortions in the reference.

\subsection{Perceptual Optimization}

Optimizing compression towards reference does not always lead to high perceptual quality of the reconstructed videos. Several prior works have explored perceptual quality optimization by mitigating artifacts or enhancing visual details with deep learning-based methods. In early works, generative adversarial networks were employed to assist in intra prediction, reducing spatial redundancy~\cite{zhu2019generative, zhang2020enhancing, zhang2021video, ramsook2022deep}, and to reconstruct high-frequency details~\cite{wang2020multi}. 
Beyond architectural or module-level enhancements, some contributions modify the optimization objectives by introducing perceptually-aware loss instead of relying solely on reference-fidelity distortion measures such as PSNR and SSIM~\cite{veerabadran2020adv_distortion,zhang2021deepvc_p,ma2024cvegan,balle2025good}. While these methods have been shown to be effective in improving the perceptual quality of PGC videos, they are neither specially designed for UGC nor tailored to address the challenges faced during transcoding.
In particular, recent work~\cite{xiong2023rate} introduces a quality saturation detection framework for UGC compression that incorporates perceptual considerations into the rate–distortion optimization process. While such approaches demonstrate the potential of perceptually guided compression, their effectiveness depends strongly on the quality of the input content.

\section{Method} 
\label{sec:method}

To address the issues mentioned above in UGC transcoding, we propose a perceptually inspired loss function, \textbf{PT-Loss}, tailored specifically for transcoding scenarios. This loss is designed to process both the non-pristine reference content fed into the transcoding codec and the reconstructed output produced by the neural codec, as illustrated in \autoref{fig:network}(A).

\subsection{Network Architecture}
\label{sec:vqa_network}

\begin{figure*}[t]
\footnotesize
\begin{minipage}[b]{0.33\linewidth}
  \centering
  \centerline{\includegraphics[width=\textwidth]{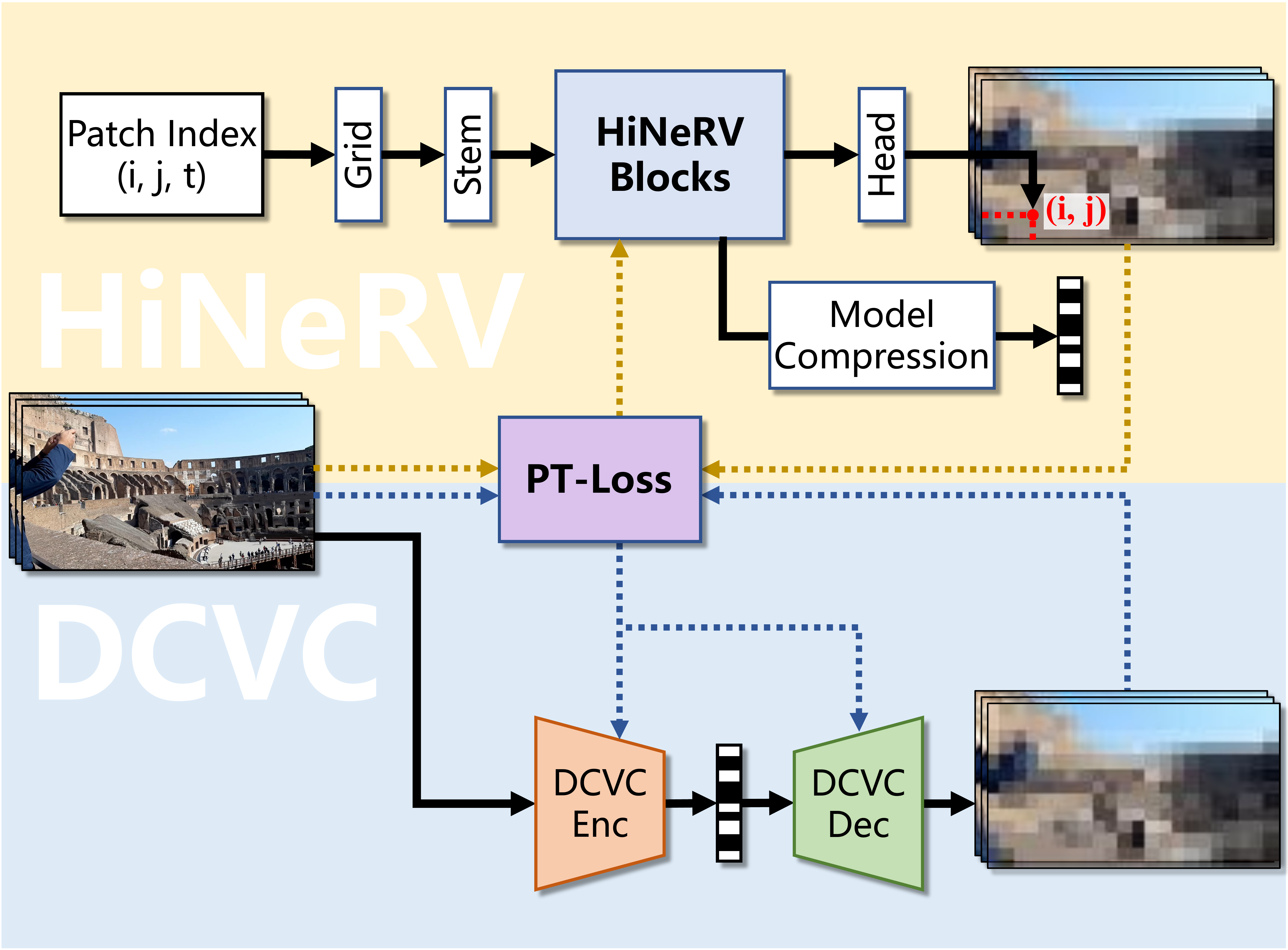}}
(A)
\label{fig2}
\end{minipage}
\begin{minipage}[b]{0.665\linewidth}
  \centering
  \centerline{\includegraphics[width=\textwidth]{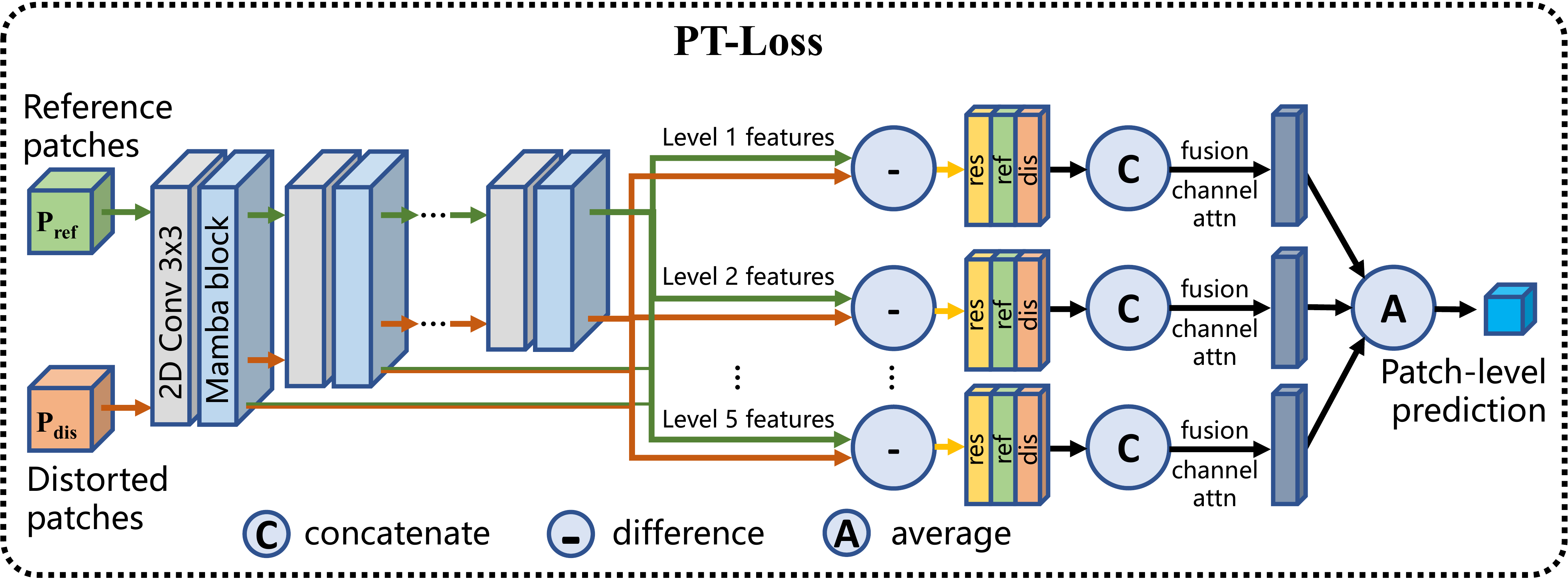}}
(B)
\label{fig3}
\end{minipage}
\caption{(A) Illustration of the PT-Loss integration into both DCVC and HiNeRV codecs. 
(B) The architecture of the PT-Loss network.}
\label{fig:network}
\vspace{-10pt}
\end{figure*}

To serve as a practical loss function, the quality model should satisfy two requirements: (i) high sensitivity to perceptual artifacts and (ii) low computational overhead to minimize training latency. The network architecture for the proposed loss function is illustrated in \autoref{fig:network}(B). Following \cite{qi2024full,feng2024rankdvqa}, the network extracts features at multiple levels from two co-located reference and distorted spatio-temporal patches, $\mathbf{R}$ and $\mathbf{D}$, with a resolution of $256 \times 256 \times 12$. In contrast to \cite{qi2024full,feng2024rankdvqa}, considering the second requirement mentioned above, we replace the prior Swin Transformer-based design~\cite{feng2024rankdvqa} with a Selective Structured State-Space Model (Mamba)~\cite{gu2024mamba} in order to efficiently learn spatio-temporal correlations by decoupling spatial feature extraction and temporal selective scanning. This enables linear complexity $O(L)$, while effectively capturing long-range temporal dependencies, which provide the capability to distinguish temporal consistency from flickering artifacts. This makes it particularly suitable for use as a perceptual loss under extended temporal context and significantly reduces inference overhead compared to attention-based designs~\cite{li2024videomamba}. The network then fuses the multi-level features and outputs a patch-level quality index. The quality indices of all the patches in the video are then aggregated into a clip-level score via a differentiable pooling layer. %Crucially, although trained via a Siamese ranking strategy (comparing pairs), the inference branch yields a deterministic, continuous scalar score, which ensures gradient stability when the metric is back-propagated through the video codec.

\subsection{Training Strategy}
\label{sec:vqa_learning}

In order to train the PT-loss function and allow it to use a distorted reference $\mathbf{R}$ as the optimization context rather than as a ground truth in the transcoding process, the loss must distinguish perceptual degradations between the absent (in practice) pristine source $\mathbf{S}$ and the distorted content $\mathbf{D}$, rather than calculating the fidelity to the non-pristine reference $\mathbf{R}$. To achieve this objective, we follow the learning strategy in \cite{qi2024full} and employ a weakly supervised, ranking-based training strategy to optimize the proposed PT-Loss. Similarly, due to the lack of large-scale human-annotated datasets developed for UGC transcoding, we produce training patch pairs while keeping their corresponding pristine sources, which are used to generate reliable quality labels using proxy metrics \cite{qi2024full}.

Specifically, distorted reference videos $\mathbf{R}$ and transcoded videos $\mathbf{D}$ are generated from pristine source videos $\mathbf{S}$ through a two-stage compression process, simulating realistic UGC delivery. The quality label of each transcoded video $\mathbf{D}$ is computed using VMAF~\cite{li2016toward} between its corresponding source $\mathbf{S}$ (rather than using $\mathbf{R}$) and $\mathbf{D}$. Although the annotations are not obtained via direct subjective scoring, their reliability has been confirmed in prior studies~\cite{feng2024rankdvqa,qi2024full} when used in a ranking-inspired learning process. Essentially, this process distills source-aware knowledge from a full-reference metric (VMAF) into a proxy network that relies solely on the non-pristine reference. In this work, a total of 252 pristine source sequences were used, extracted from the BVI-DVC dataset~\cite{ma2021bvi}, the CVPR 2022 CLIC challenge training dataset~\cite{clic2022} and the YouTube-UGC 2K database~\cite{wang2019youtube} alongside six self-captured videos to further extend UGC content diversity. These source sequences $\mathbf{S}$ were first compressed using the fast implementation of H.264/AVC ~\cite{h264_overview}, x264\footnote{It is noted that we only used conventional codecs to produce the training material due to their fast coding speeds. Further improvement may be achieved if neural codecs are used here, but this remains future work.}~\cite{x264repos} with QP= 30, 37, 42 to produce non-pristine references $\mathbf{R}$. The latter was then transcoded using three different codecs (x264, AV1, VP9~\cite{vp9}) with quantization parameters 32--42 (x264) with an interval of 2 and 38--63 (AV1, VP9) with an interval of 5.  

The generated reference and distorted video pairs are then randomly cropped into $256 \times 256 \times 12$ patch pairs, which are annotated using the method mentioned above. This results in a total of 604,800 patch pairs that are used to train the proposed PT-Loss via a Siamese ranking-based learning strategy proposed in \cite{feng2024rankdvqa,qi2024bvi}. During training, we sampled single-source and cross-source patch pairs with an 8:2 ratio to encourage the model to learn robust and content-invariant perceptual features. After convergence, the model is further fine-tuned using only patch pairs from the single source, as accurately modeling quality differences for the same content is more critical than cross-content ranking performance for the loss function.

\subsection{Integration into Neural Video Codecs}
\label{sec:codec_integration}

To validate the effectiveness of the proposed loss function, we integrated it into two different neural video codecs as the basis for rate-quality optimization, including an autoencoder-based codec~\cite{li2021deep} and an overfitting (or implicit neural representation based) codec, HiNeRV~\cite{kwan2023hinerv}, as shown in \autoref{fig:network}(A).

\begin{table*}[t]
\centering
\caption{BD-rate (measured in PSNR/VMAF) results on the BVI-UGC dataset across different reference groups. Negative values indicate bitrate savings compared to the baseline. The complexity is computed by taking the original video codecs as baselines. Since HiNeRV contains multiple scales of models, its complexities are presented as \textit{Smallest--Largest} models.}
\resizebox{\linewidth}{!}{
\begin{tabular}{r|rr|rr|rr|rr|rrr}
\toprule
\multirow{2}{*}{} 
& \multicolumn{2}{c|}{\textbf{\textcolor{red!85!black}{Low}}} 
& \multicolumn{2}{c|}{\textbf{\textcolor{yellow!65!black}{Medium}}} 
& \multicolumn{2}{c|}{\textbf{\textcolor{green!70!black}{High}}}
& \multicolumn{2}{c|}{\textbf{Overall}} 
& \multicolumn{3}{c}{\textbf{Complexity}} \\
\cmidrule{2-12}
& \textbf{PSNR} & \textbf{VMAF} 
& \textbf{PSNR} & \textbf{VMAF} 
& \textbf{PSNR} & \textbf{VMAF} 
& \textbf{PSNR} & \textbf{VMAF} 
& MACs & Params & EncT \\
\midrule
DCVC &  0.00\%  &  0.00\%  &  0.00\%  &  0.00\%  &  0.00\% &  0.00\% &  0.00\% &  0.00\% & 2268G & 7.94M & 100.0\%  \\
+\cite{qi2024full} &  -8.54\%  &  -12.19\%  &  -7.22\%  &  -7.97\%  &  \textbf{-7.91\%}  &  -6.29\%  &  -7.90\%  &  -7.93\%  & +0.0G & +0.0M & 100.0\%  \\
\midrule
\textbf{+PT-Loss (ours)} &  \textbf{-12.98\%}  &  \textbf{-12.49\%}  &  \textbf{-8.11\%}  &  \textbf{-9.21\%}  &  -7.84\%  &  \textbf{-6.93\%}  & \textbf{-8.46\%} &  \textbf{-8.64\%}  & +0.0G & +0.0M & 100.0\%   \\
\midrule
\midrule
HiNeRV &  0.00\%  &  0.00\%  &  0.00\%  &  0.00\%  &  0.00\% &  0.00\% &  0.00\% &  0.00\% & 43--718G & 0.76--12.8M & 100.0\%  \\
+\cite{qi2024full} &  -15.95\%  &  -16.18\%  &  \textbf{-13.46\%}  &  -11.67\%  &  -10.51\%  &  -7.76\%  &  -12.70\%  &  -10.74\% & +5.2G & +4.6M & 195.0\%  \\
\midrule
\textbf{+PT-Loss (ours)}   &  \textbf{-16.15\%}  &  \textbf{-16.46\%}  &  -13.44\%  &  \textbf{-11.72\%}  &  \textbf{-10.65\%}  &  \textbf{-7.83\%}  &  \textbf{-12.89\%}  &  \textbf{-10.83\%} & +1.4G & +0.8M & 121.3\%  \\
\bottomrule
\end{tabular}}
\label{tab:bdrate_results}
\vspace{-5pt}
\end{table*}

Specifically, DCVC adopts a learning-based inter-frame predictive coding paradigm, which employs a pre-trained image codec for the independent spatial compression of anchor frames and a specialized P-frame model that leverages inter-frame correlations. To integrate the proposed perceptual loss, we focus exclusively on fine-tuning the P-frame model on the training split of the Vimeo-90k septuplet dataset~\cite{xue2019video}, while keeping other modules fixed. The total loss is a combination of the original MSE loss and the proposed PT-Loss, $\mathcal{L}_\mathrm{PT}$. 
\begin{equation}
    \mathcal{L}_{total} =  (1-\alpha)  \mathcal{L}_{MSE} +\alpha \mathcal{L}_\mathrm{PT}.
\end{equation}
Here $\alpha$ is a hyperparameter to balance $\mathcal{L}_{MSE}$ and $\mathcal{L}_\mathrm{PT}$. To address the magnitude discrepancy between the different objectives, we align the magnitude of $\mathcal{L}_\mathrm{PT}$ to that of $\mathcal{L}_{MSE}$ adaptively by calculating the ratio of their mean values in the first five training iterations. The weighting coefficient $\alpha$ is then applied to balance their contributions.

In contrast to DCVC, HiNeRV is an implicit neural representation (INR) based video codec that represents an entire video as a continuous neural representation and performs compression by overfitting the network parameters to a target video sequence. During the per-video overfitting, the original HiNeRV optimizes a combined loss consisting of $\ell_1$ and MS-SSIM. In our integration, we replaced the MS-SSIM loss with the proposed PT-Loss, as given below.
 \begin{equation}
 \mathcal{L}_{total} = (1-\alpha)\mathcal{L}_{1} +  \alpha \mathcal{L}_\mathrm{PT}.
 \label{eq:lossHiNeRV}
 \end{equation}

\section{Results and Discussion}

In order to evaluate the two neural video codecs in conjunction with the proposed loss function, we conducted an experiment using the BVI-UGC dataset~\cite{qi2024bvi}, which is specifically designed for UGC video transcoding. The dataset contains 60 uncompressed (HD) source videos spanning 15 common UGC categories, covering a wide range of real-world content characteristics, including various artifacts, motion blur, defocus, over- and under-exposure, screen content, as well as both landscape and portrait formats. For each source sequence, the database also contains three non-pristine reference sequences (compressed by x264) at various quantization levels, QP = 30, 37, and 42, simulating high-, medium-, and low-quality reference conditions, respectively. 

Due to the high computational cost associated with video compression, in this experiment, we selected a subset of 15 source sequences from BVI-UGC, following the selection strategy proposed in~\cite{zhang2018bvi} to ensure wide feature coverage and a uniform feature distribution across the database~\cite{winkler2012analysis}, with one representative video from each UGC category. BVI-UGC is chosen because it provides high-quality source videos and has been previously used in research on UGC transcoding~\cite{qi2024bvi}. Other available UGC datasets~\cite{hosu2017konstanz, sinno2018large, yu2021predicting, li2020ugc,wang2019youtube} do not provide high-quality source videos.
The corresponding non-pristine references are used as the input to the transcoding process when using the tested video codecs. This results in three groups of reference sequences, each containing 15 videos. When measuring the video quality of the transcoded content, instead of using the non-pristine references, the uncompressed source videos are employed as the anchor in order to ensure a reliable quality evaluation. 
%This is one of the major reasons that BVI-UGC is used in this experiment, considering other available UGC databases~\cite{hosu2017konstanz, sinno2018large, wang2019youtube, yu2021predicting, li2020ugc} only contain non-pristine references (no ground truth sources).

We compared the integrated video codecs, annotated as DCVC+PT-Loss and HiNeRV+PT-Loss, with their original counterparts (based on their original training losses), DCVC and HiNeRV. For both codecs, the weighting parameter, $\alpha$, is set to 0.2. To measure the video quality, we employ two quality metrics, PSNR and VMAF -- the former focuses on the pixel-wise differences, while the latter is a perceptual metric that correlates well with human perception. As mentioned above, both quality metrics are calculated between the transcoded content and its corresponding uncompressed source (rather than the input non-pristine reference for compression) to ensure the reliability of the quality prediction (see \autoref{fig:transcoding}).

\subsection{Quantitative Results}

\begin{figure*}[!t]
\centering
\small
\begin{minipage}[c]{\linewidth}
\centering
    \begin{minipage}[c]{0.24\linewidth}
    \centering
    \centerline{\includegraphics[width=1.05\columnwidth]{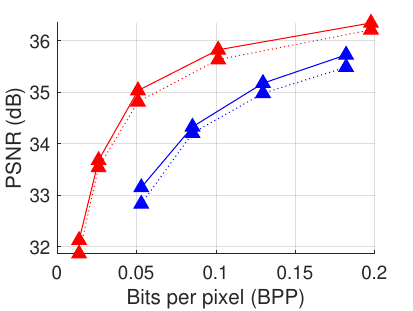}}
    high
    \end{minipage}
    \begin{minipage}[c]{0.24\linewidth}
    \centering
    \centerline{\includegraphics[width=1.05\columnwidth]{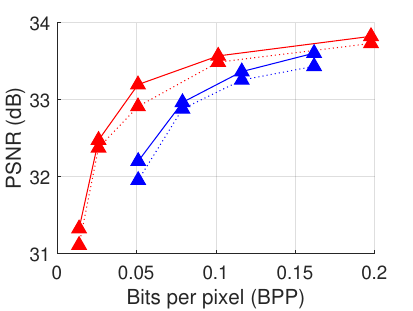}}
    medium
    \end{minipage}
    \begin{minipage}[c]{0.24\linewidth}
    \centering
    \centerline{\includegraphics[width=1.05\columnwidth]{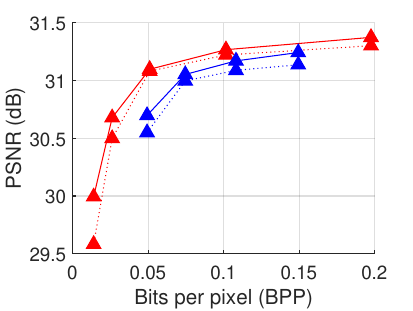}}
    low
    \end{minipage}
    \begin{minipage}[c]{0.24\linewidth}
    \centering
    \centerline{\includegraphics[width=1.05\columnwidth]{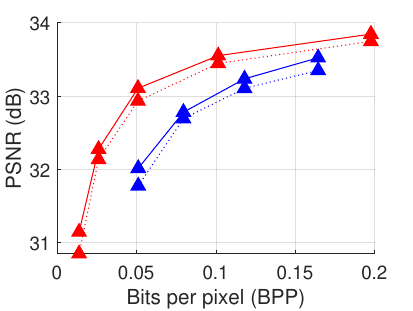}}
    overall
    \end{minipage}

    \begin{minipage}[c]{0.24\linewidth}
    \centering
    \centerline{\includegraphics[width=1.05\columnwidth]{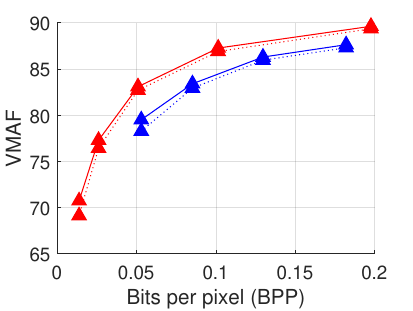}}
    \end{minipage}
    \begin{minipage}[c]{0.24\linewidth}
    \centering
    \centerline{\includegraphics[width=1.05\columnwidth]{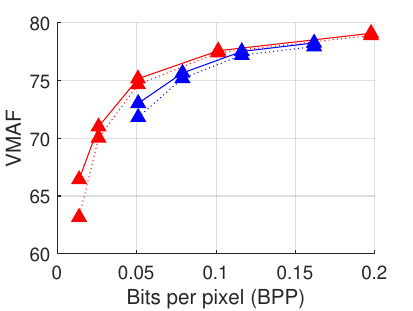}}
    \end{minipage}
    \begin{minipage}[c]{0.24\linewidth}
    \centering
    \centerline{\includegraphics[width=1.05\columnwidth]{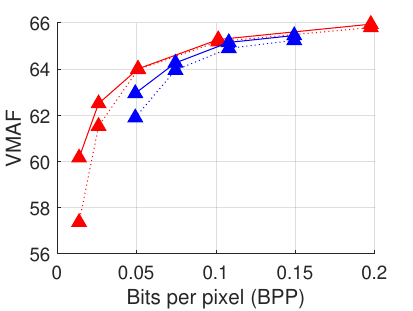}}
    \end{minipage}
    \begin{minipage}[c]{0.24\linewidth}
    \centering
    \centerline{\includegraphics[width=1.05\columnwidth]{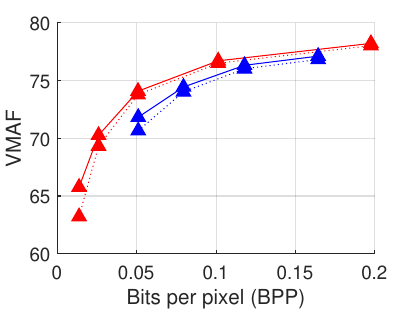}}
    \end{minipage}
    
    \begin{minipage}[c]{0.58\linewidth}
    \centering
    \centerline{\includegraphics[width=1.05\columnwidth]{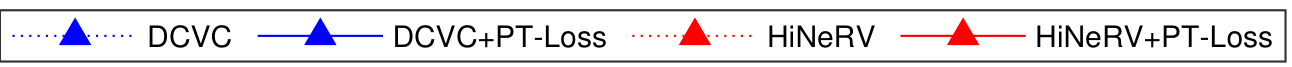}}
    \end{minipage}
\end{minipage}
\caption{Overall and group-wise compression results on BVI-UGC datasets across different reference groups.} 
\label{fig_bdrate}
\vspace{-5pt}
\end{figure*}

\begin{figure*}[t]
\small
\centering
    \begin{minipage}[c]{0.24\linewidth}
    \centering
    \centerline{\includegraphics[width=\columnwidth]{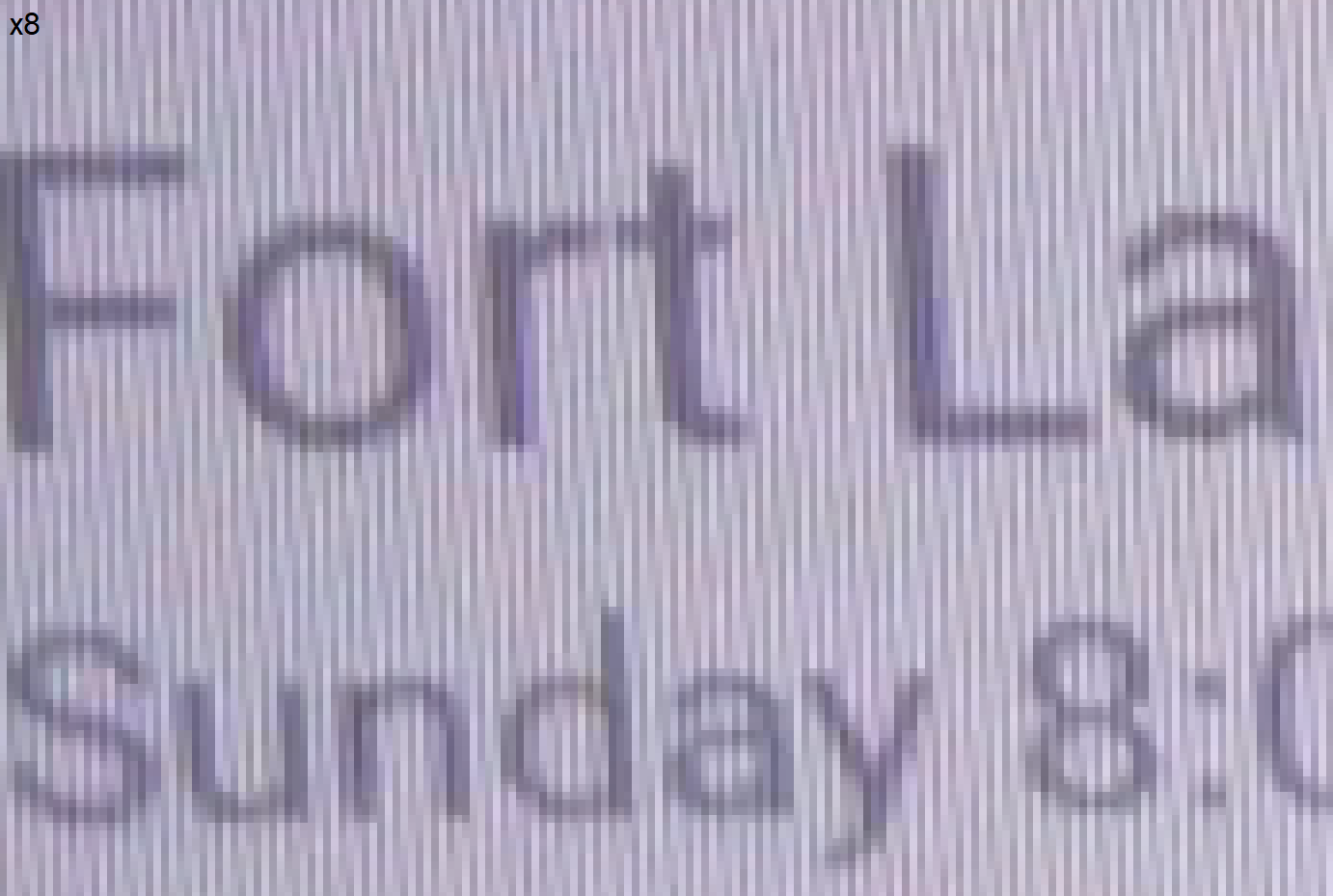}}
    \textbf{original}
    \end{minipage}
    \begin{minipage}[c]{0.24\linewidth}
    \centering
    \centerline{\includegraphics[width=\columnwidth]{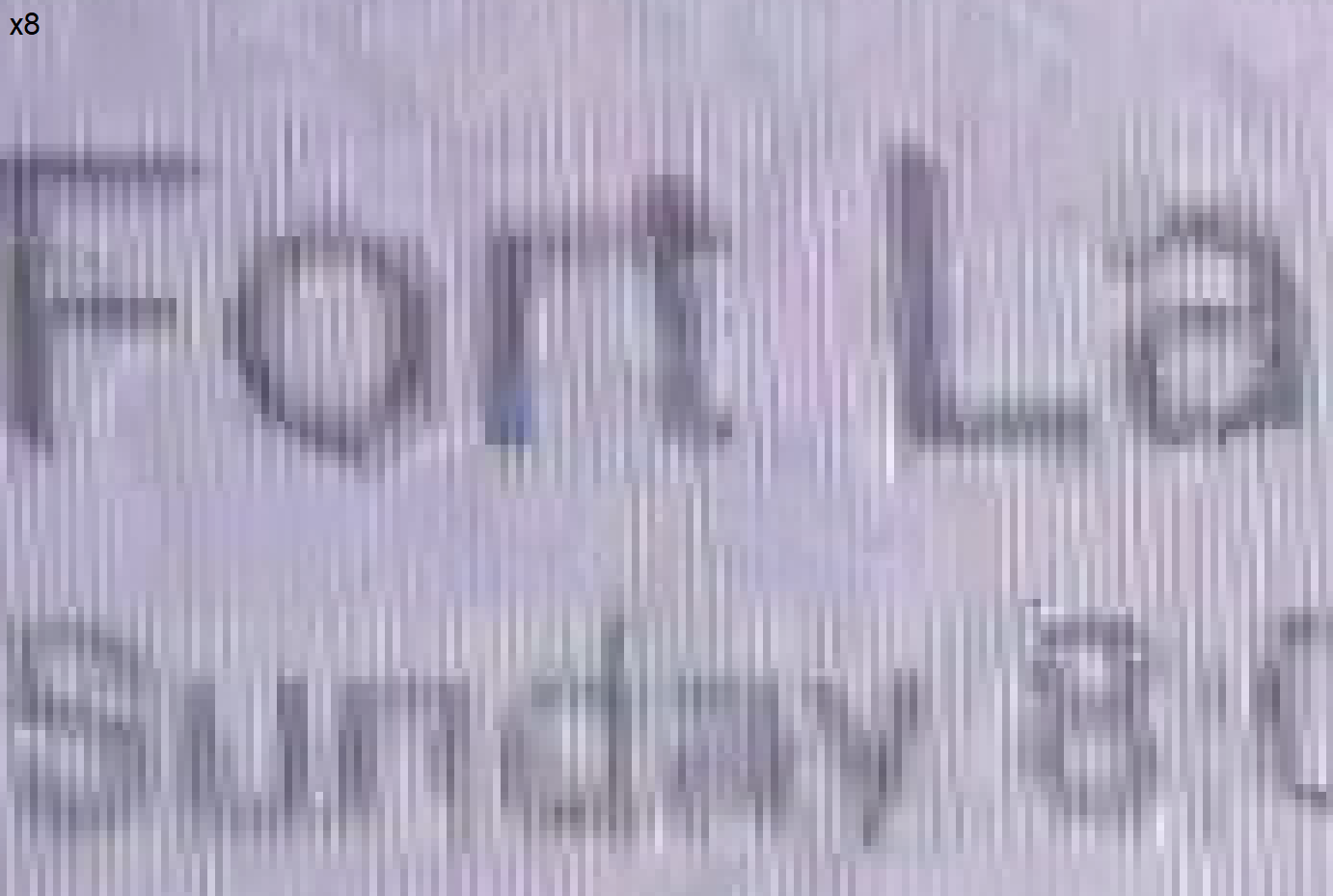}}
    reference QP42 \textbf{(33.96dB)}
    \end{minipage}
    \begin{minipage}[c]{0.24\linewidth}
    \centering
    \centerline{\includegraphics[width=\columnwidth]{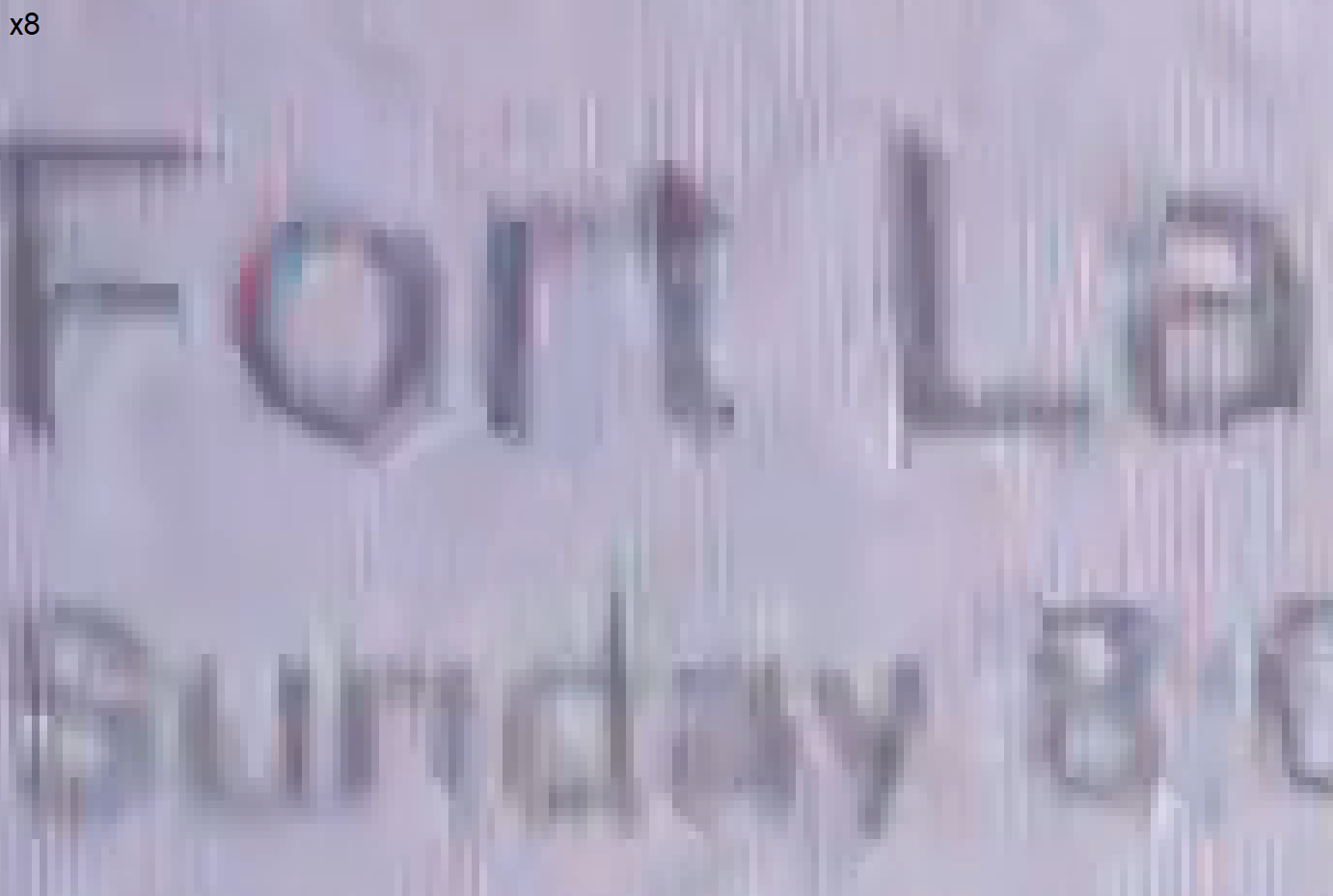}}
    DCVC bpp=0.1701 \textbf{(33.23dB)}
    \end{minipage}
    \begin{minipage}[c]{0.24\linewidth}
    \centering
    \centerline{\includegraphics[width=\columnwidth]{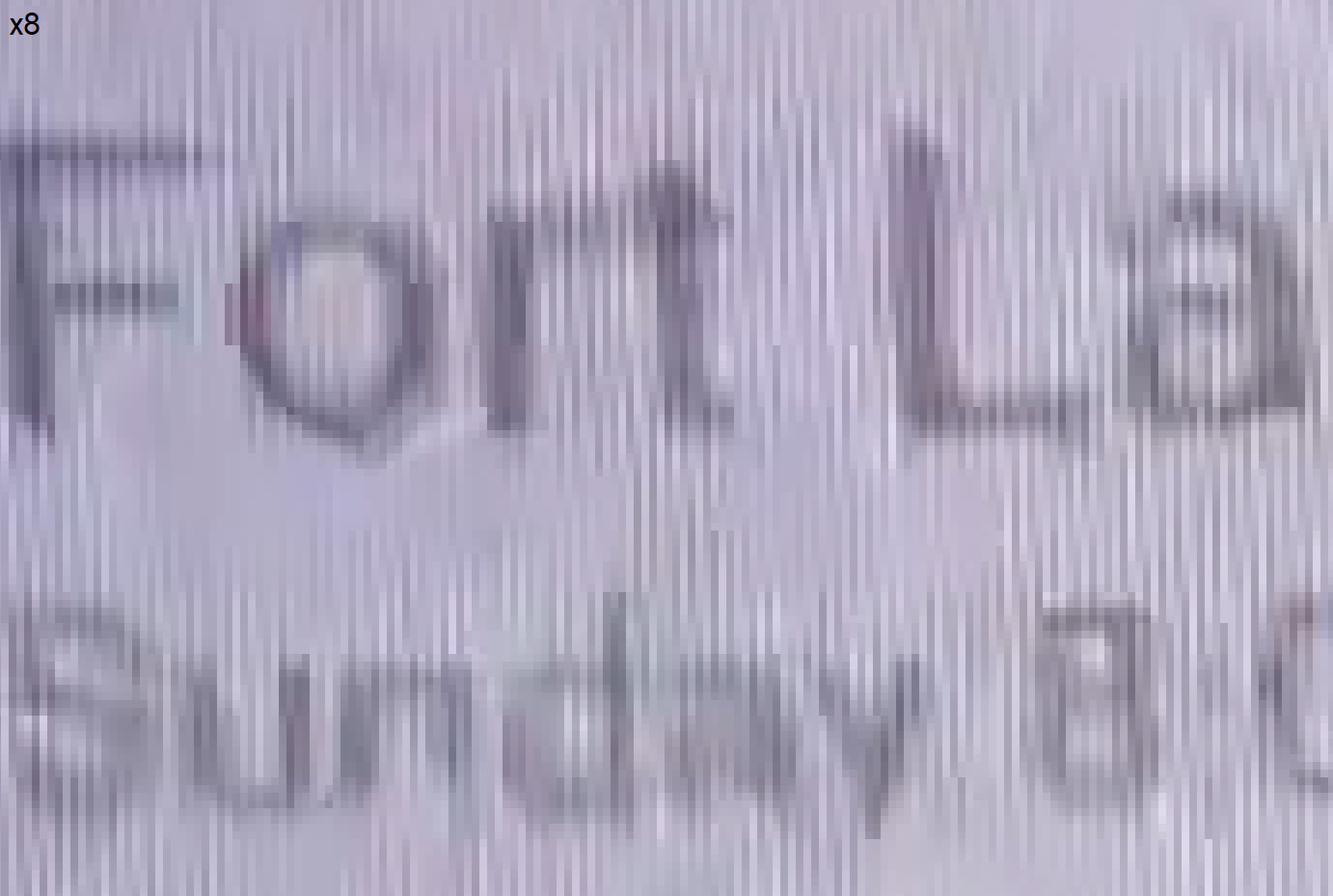}}
    +PT-Loss bpp=0.1681 \textbf{(34.33dB)}
    \end{minipage}

    \begin{minipage}[c]{0.24\linewidth}
    \centering
    \centerline{\includegraphics[width=\columnwidth]{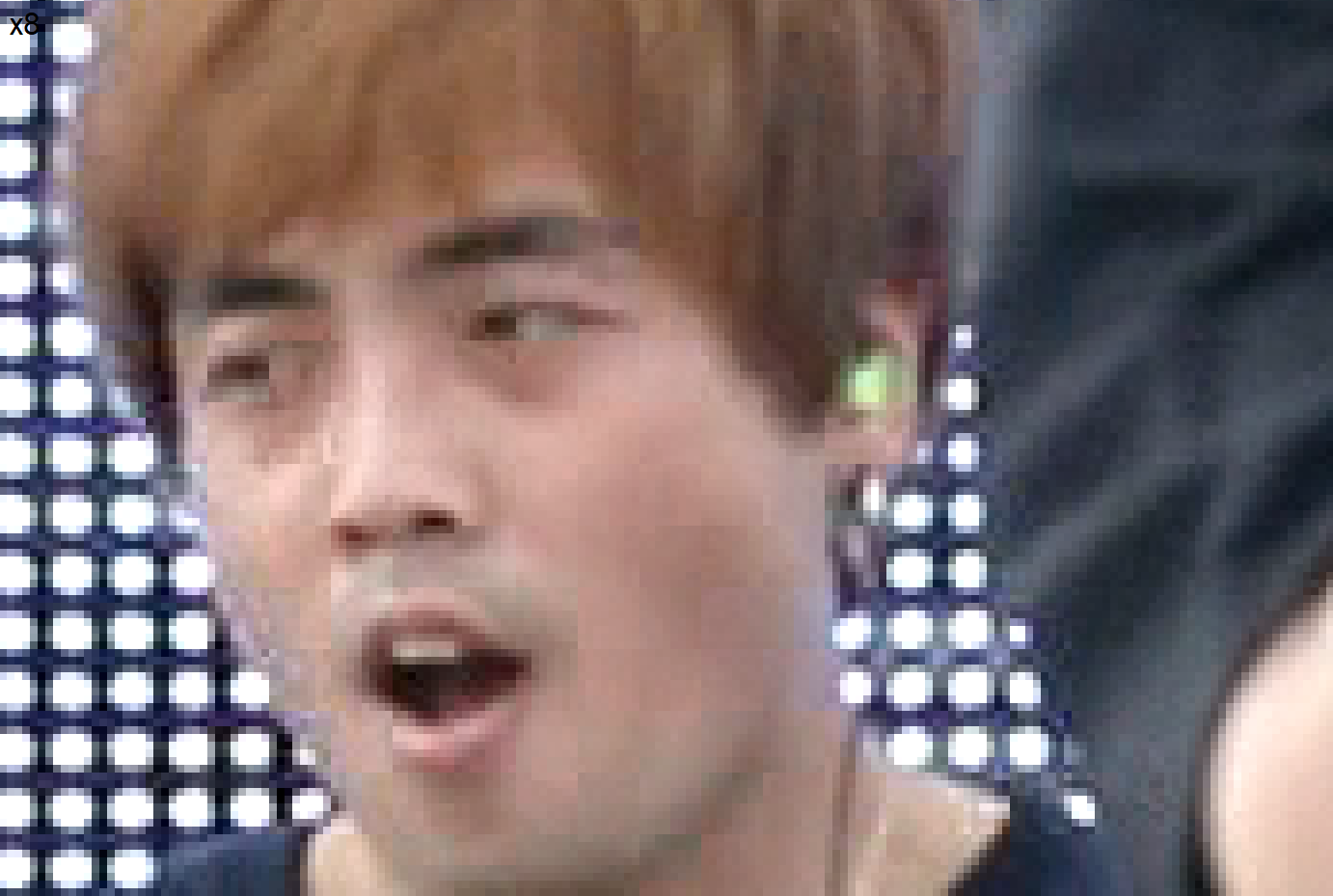}}
    \textbf{original}
    \end{minipage}
    \begin{minipage}[c]{0.24\linewidth}
    \centering
    \centerline{\includegraphics[width=\columnwidth]{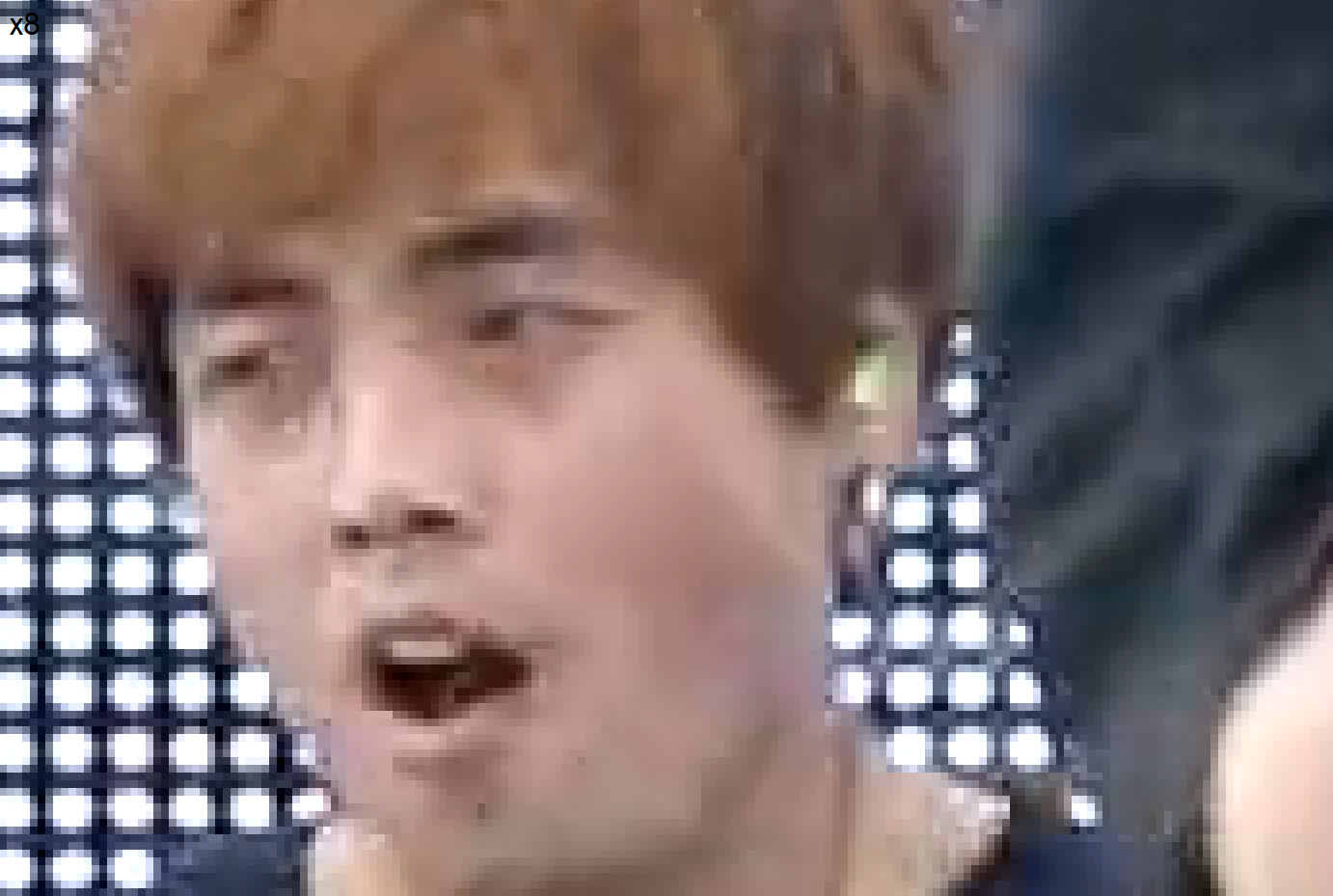}}
    reference QP37 \textbf{(32.56dB)}
    \end{minipage}
    \begin{minipage}[c]{0.24\linewidth}
    \centering
    \centerline{\includegraphics[width=\columnwidth]{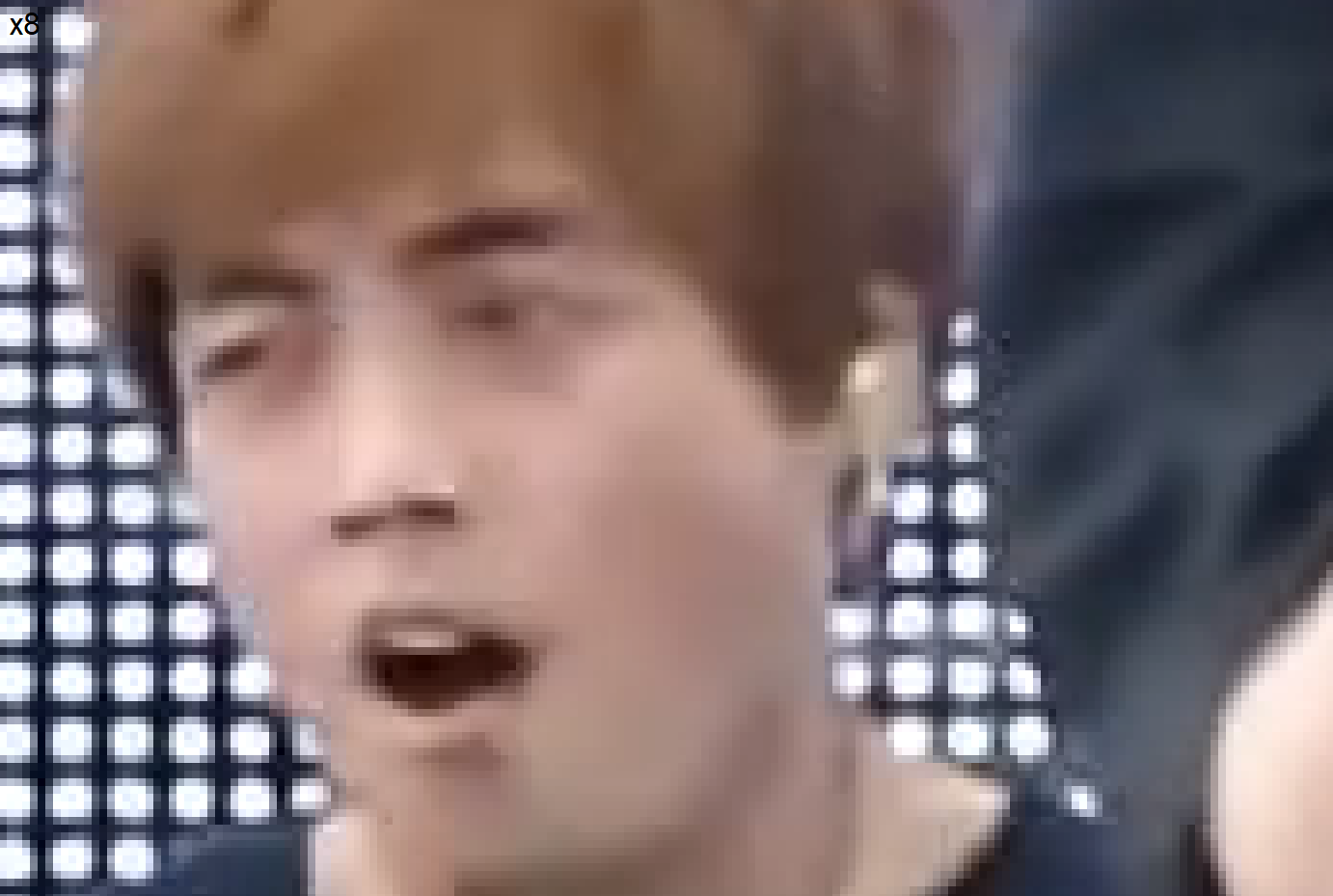}}
    DCVC bpp=0.1165 \textbf{(26.53dB)}
    \end{minipage}
    \begin{minipage}[c]{0.24\linewidth}
    \centering
    \centerline{\includegraphics[width=\columnwidth]{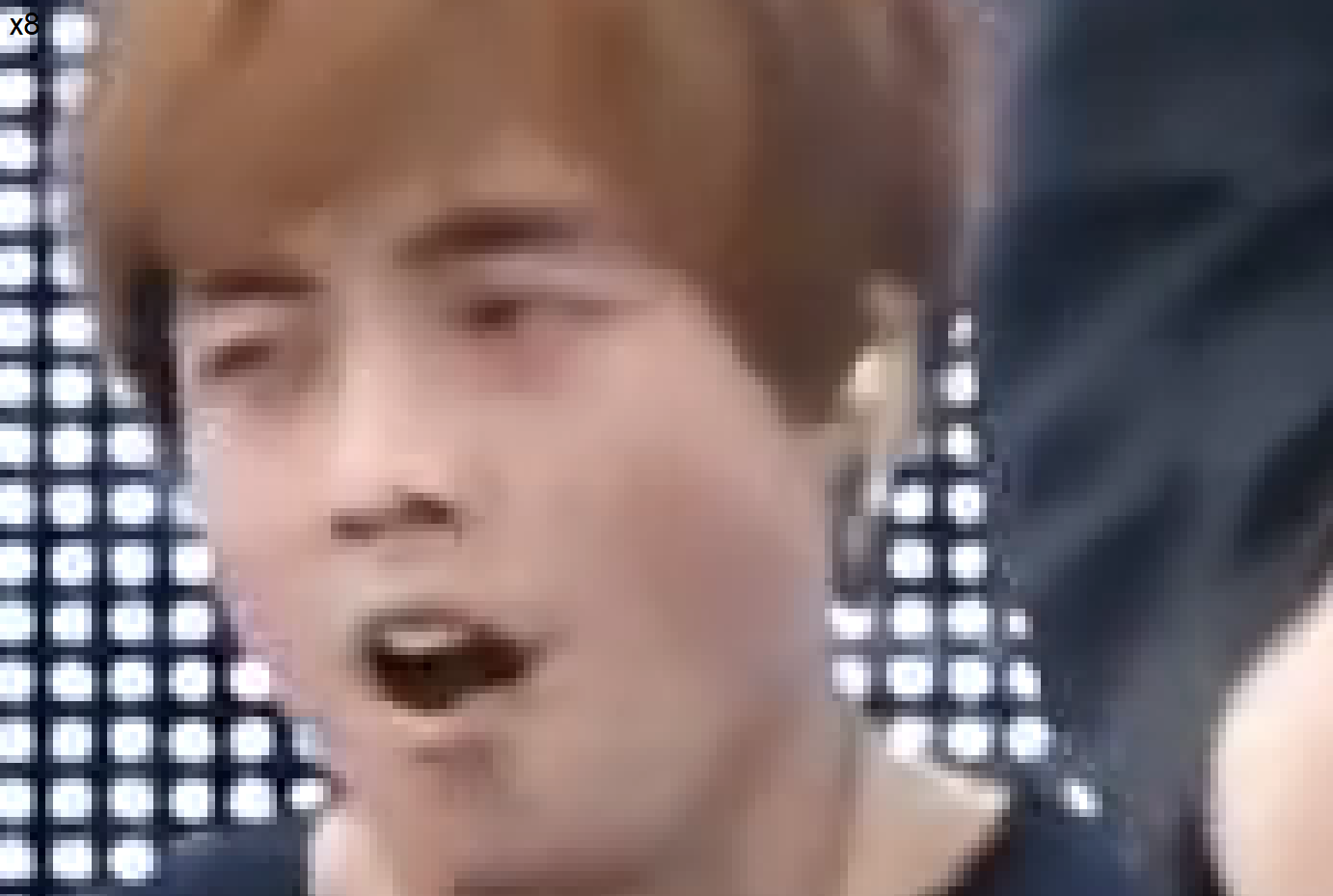}}
    +PT-Loss bpp=0.1139 \textbf{(26.66dB)}
    \end{minipage}

\caption{Visual comparison between DCVC and DCVC+PT-Loss. } 
\label{fig_viscompare1}
\vspace{-10pt}
\end{figure*}

\autoref{tab:bdrate_results} summarizes the average BD-rate results between each integrated codec and its original baseline, with the quality measured in PSNR and VMAF for all 15 tested source sequences. It can be observed that by incorporating the proposed loss, significant overall coding gains have been achieved, with 8.46\% (in PSNR) and 8.64\% (in VMAF) BD-rate savings over DCVC, and 12.89\% and 10.83\% over HiNeRV. Specifically, the performance improvement is more evident when the quality of the input reference content is lower. This observation is consistent with the motivation of this work. It is also noted that, when the PT-Loss is integrated into HiNeRV, more significant coding gains are possible compared to DCVC. This may be due to the overfitting nature of the INR-based codecs. To further verify coding gains across the full bitrate range, we provide the rate–distortion (R–D) curves for both anchor codecs on the whole test database and the three reference groups, as shown in \autoref{fig_bdrate}. It can be observed that the proposed loss function achieves consistent coding gains in most cases. Notably, these improvements are more pronounced in the low-quality reference groups.

% \subsection{Complexity Analysis} 

As mentioned in \autoref{sec:vqa_network}, one of the primary requirements for a loss function is low computational complexity. In \autoref{tab:bdrate_results}, we provide the complexity figures of encoders with and without the proposed PT-Loss, in terms of parameter count, MACs and relative encoding time (to the corresponding baselines). We did not include the decoding complexity figures, as the loss function only applies in the encoding process, while the decoding complexity remains unchanged for both codecs. For the DCVC codec, as the perceptual loss is exclusively utilized as a training objective to supervise the offline optimization of the encoder and decoder, it does not contribute to the online inference process. Therefore, the encoding complexity also remains the same. For HiNeRV, due to the overfitting nature of the codec, the new loss function is used in the encoding (training) process, introducing a 21.3\% encoding time overhead, a 0.8M parameter count increase, and 1.4G additional MACs. 

\subsection{Qualitative Results} 

\begin{figure*}[t]
\centering
\small
\centering
    \begin{minipage}[c]{0.24\linewidth}
    \centering
    \centerline{\includegraphics[width=\columnwidth]{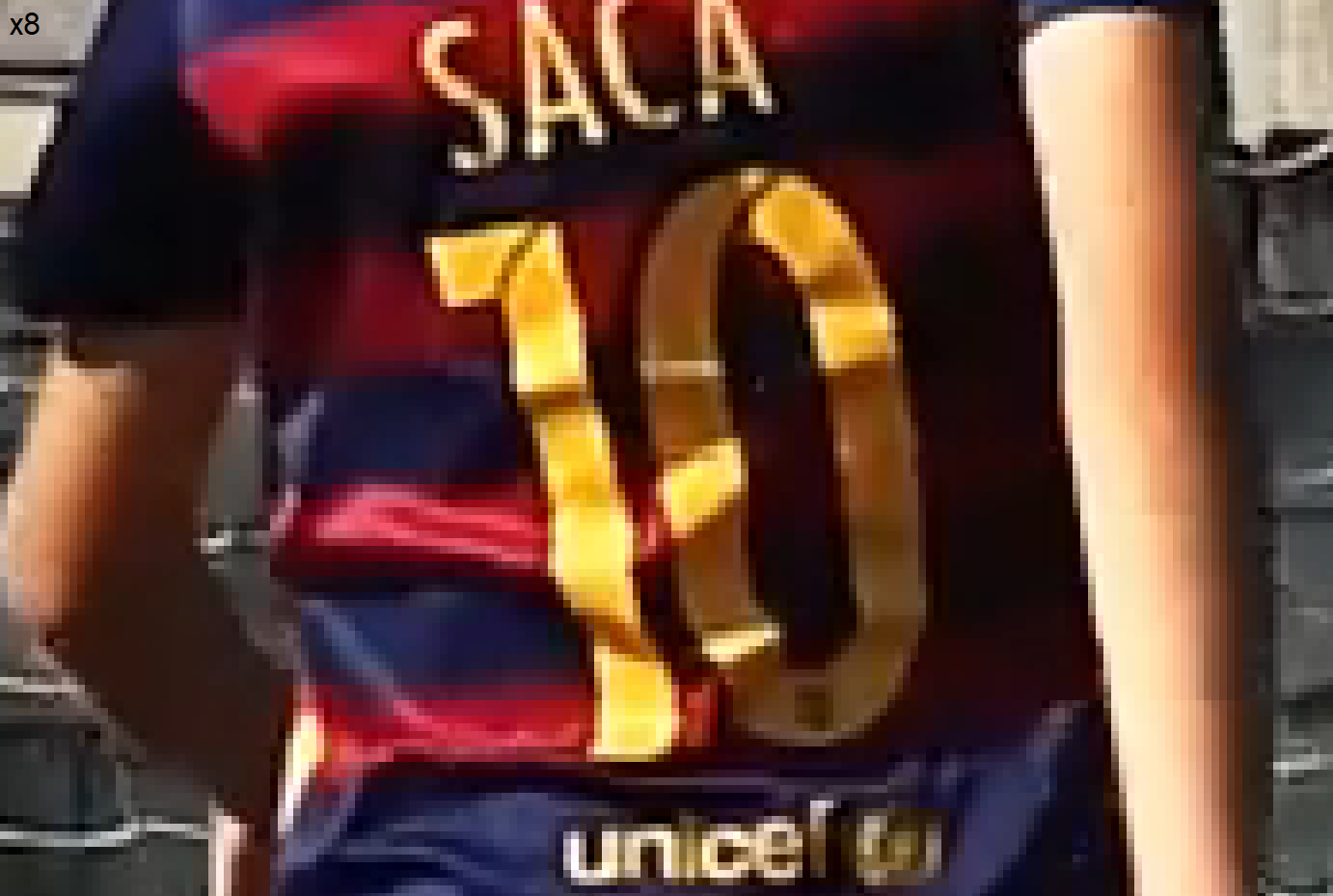}}
    \textbf{original}
    \end{minipage}
    \begin{minipage}[c]{0.24\linewidth}
    \centering
    \centerline{\includegraphics[width=\columnwidth]{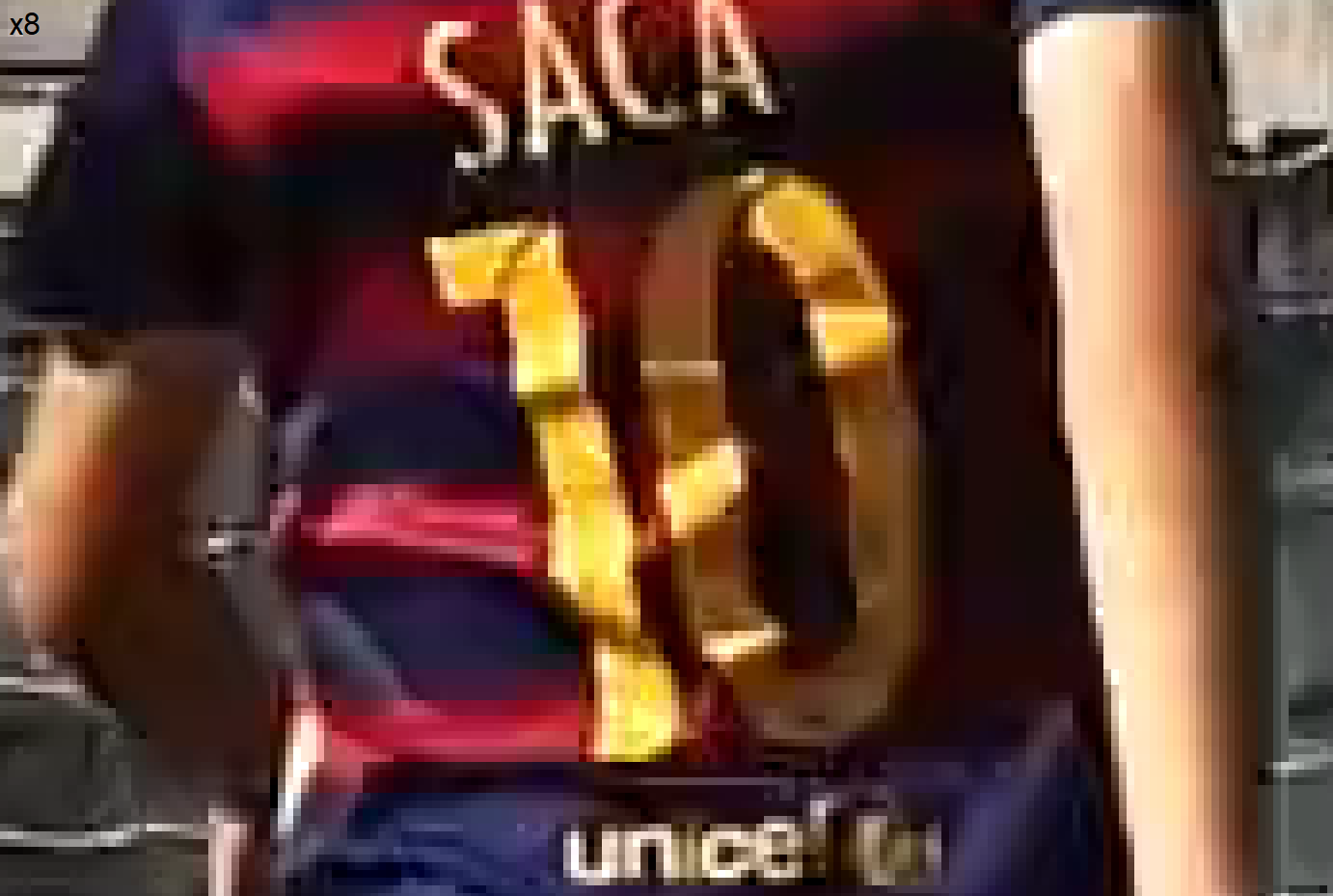}}
    reference QP42 \textbf{(24.35dB)}
    \end{minipage}
    \begin{minipage}[c]{0.24\linewidth}
    \centering
    \centerline{\includegraphics[width=\columnwidth]{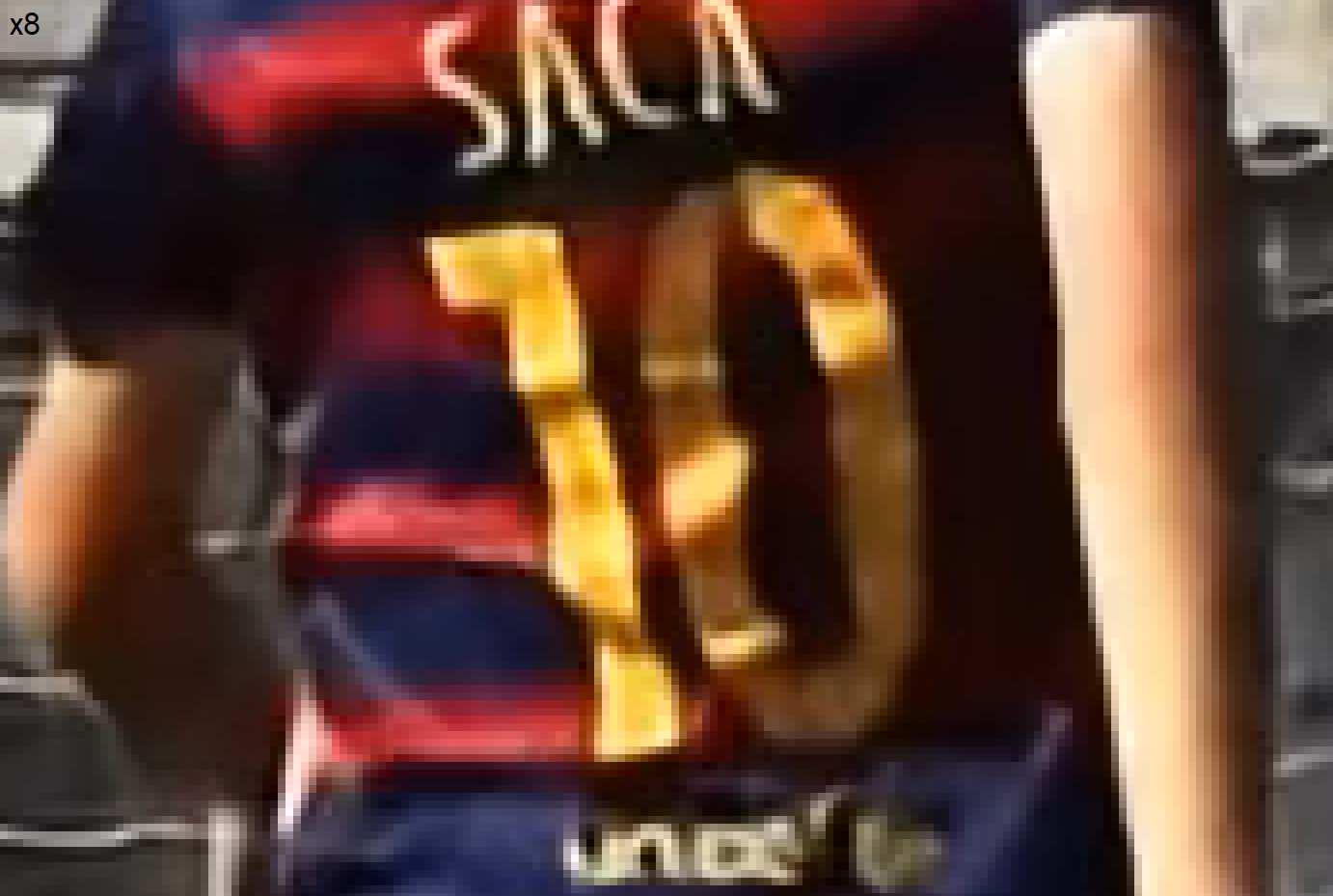}}
    HiNeRV bpp=0.0999 \textbf{(23.06dB)}
    \end{minipage}
    \begin{minipage}[c]{0.24\linewidth}
    \centering
    \centerline{\includegraphics[width=\columnwidth]{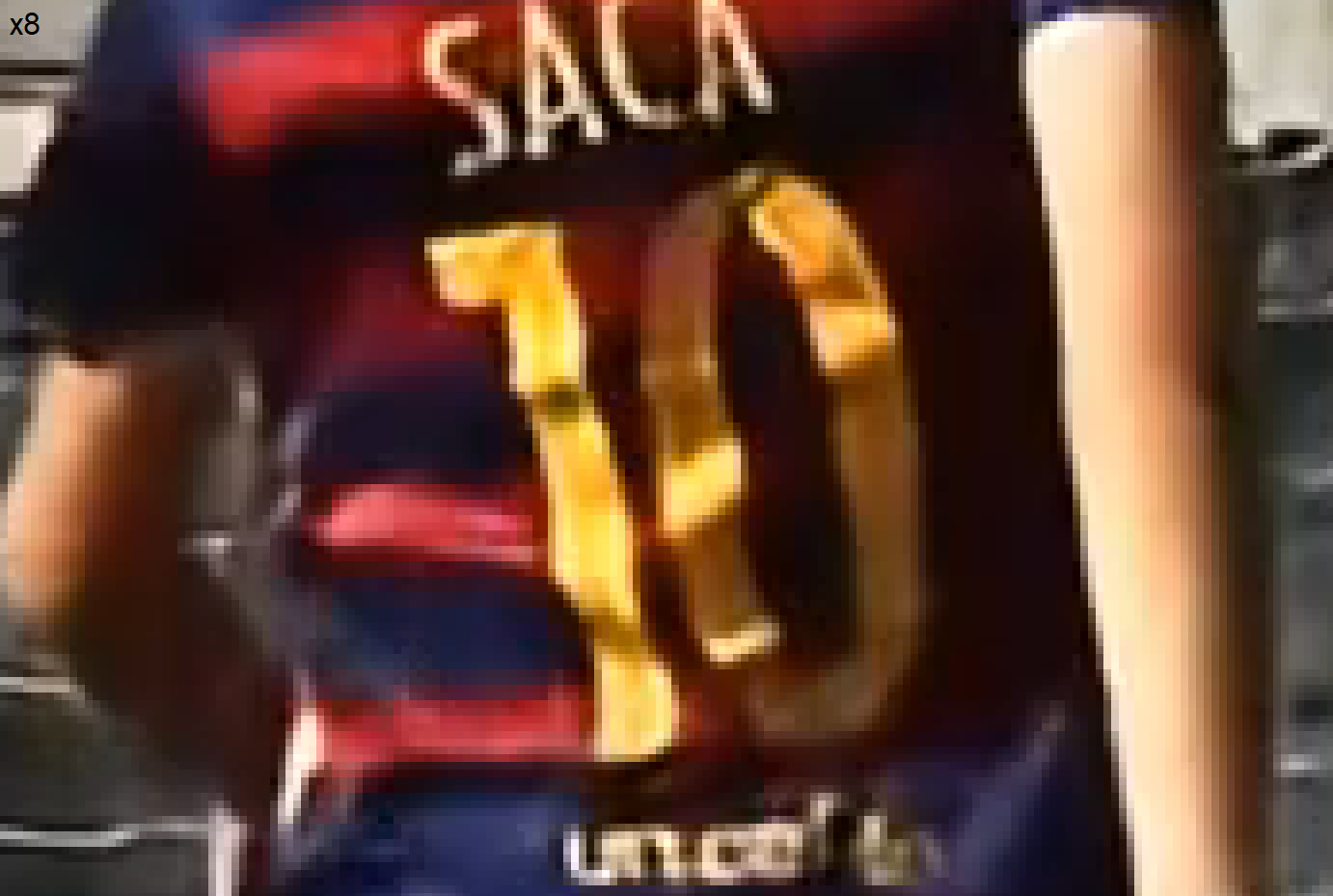}}
    +PT-Loss bpp=0.1001 \textbf{(23.83dB)}
    \end{minipage}

\centering
    \begin{minipage}[c]{0.24\linewidth}
    \centering
    \centerline{\includegraphics[width=\columnwidth]{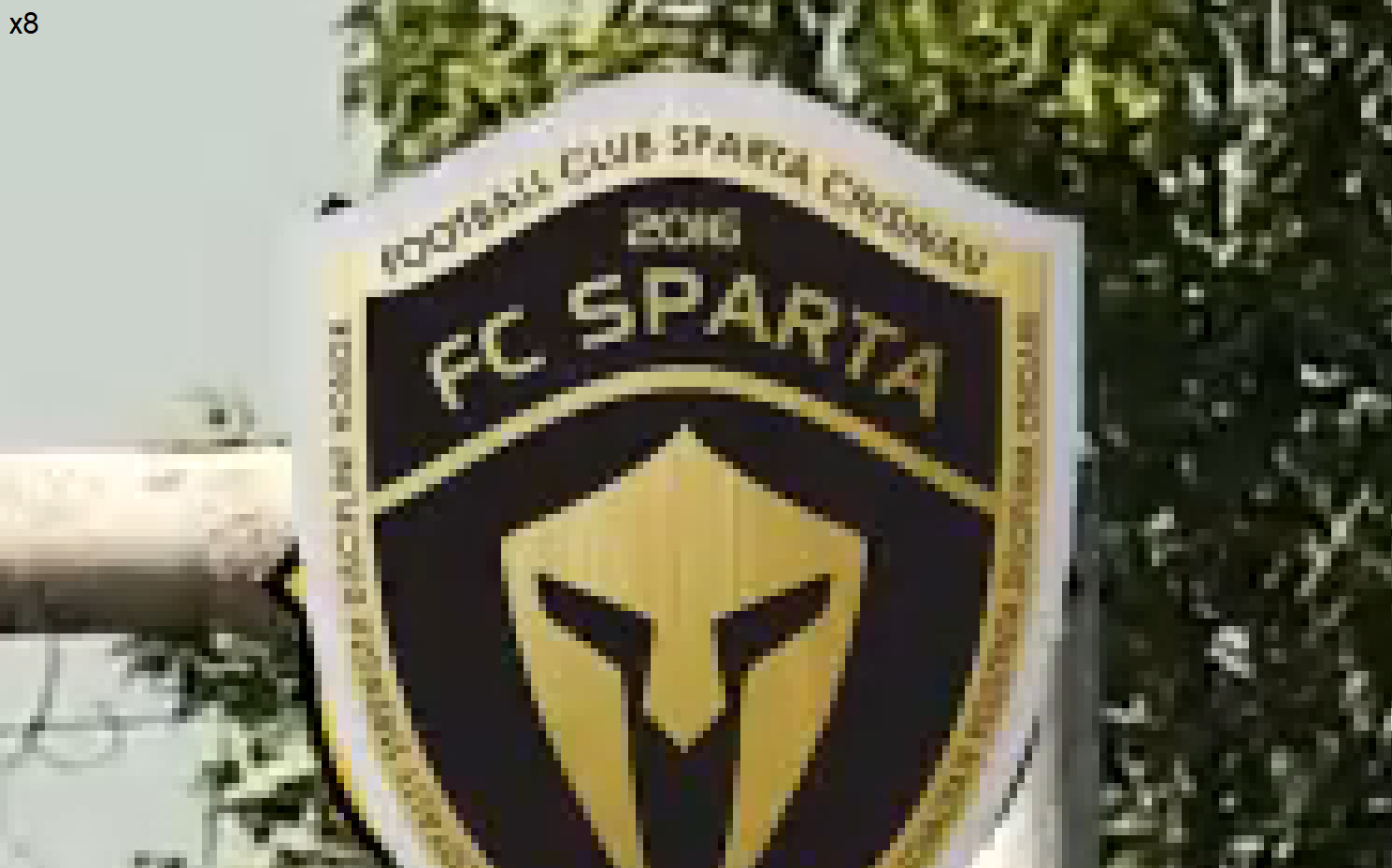}}
    \textbf{original}
    \end{minipage}
    \begin{minipage}[c]{0.24\linewidth}
    \centering
    \centerline{\includegraphics[width=\columnwidth]{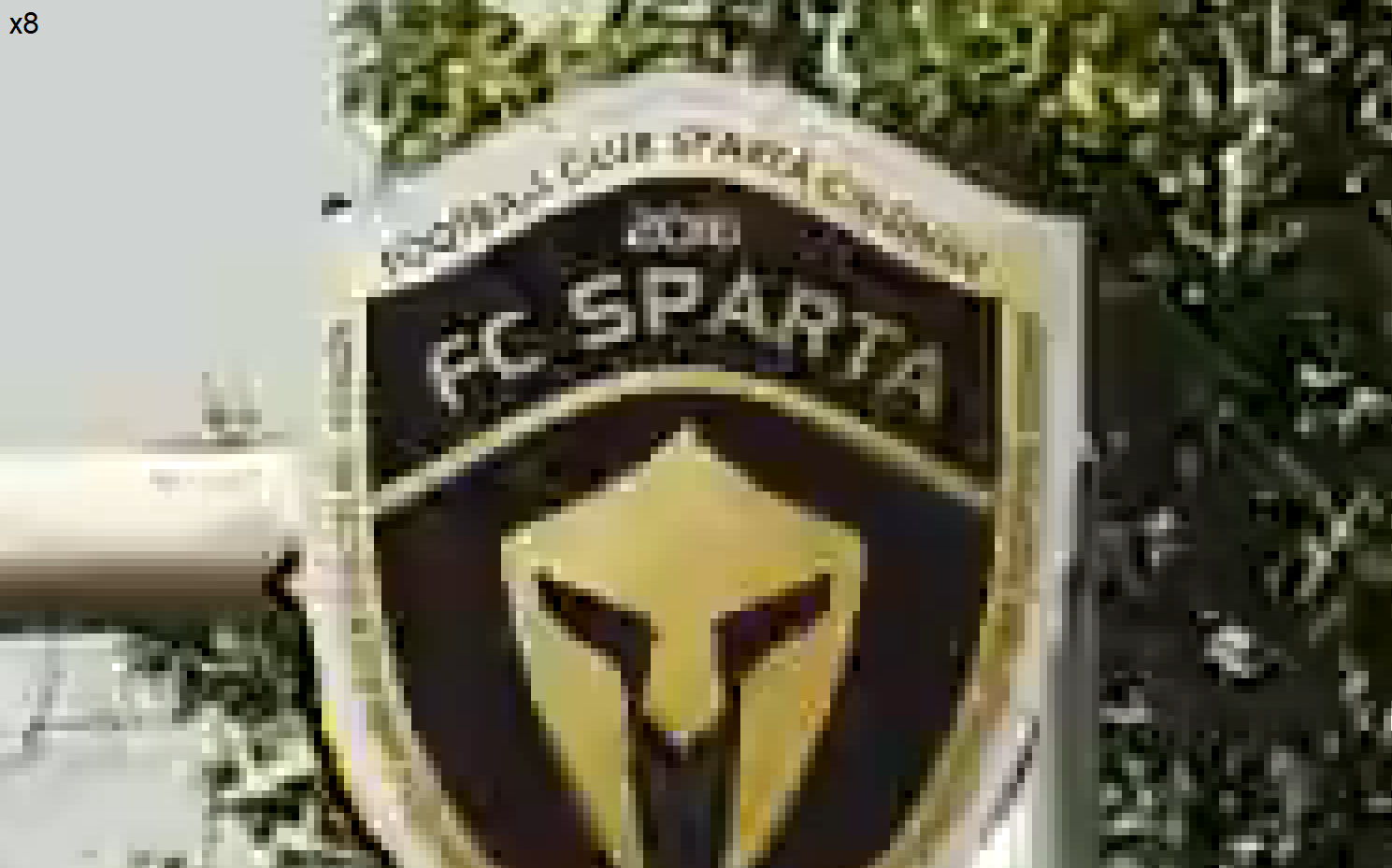}}
    reference QP42 \textbf{(24.35dB)}
    \end{minipage}
    \begin{minipage}[c]{0.24\linewidth}
    \centering
    \centerline{\includegraphics[width=\columnwidth]{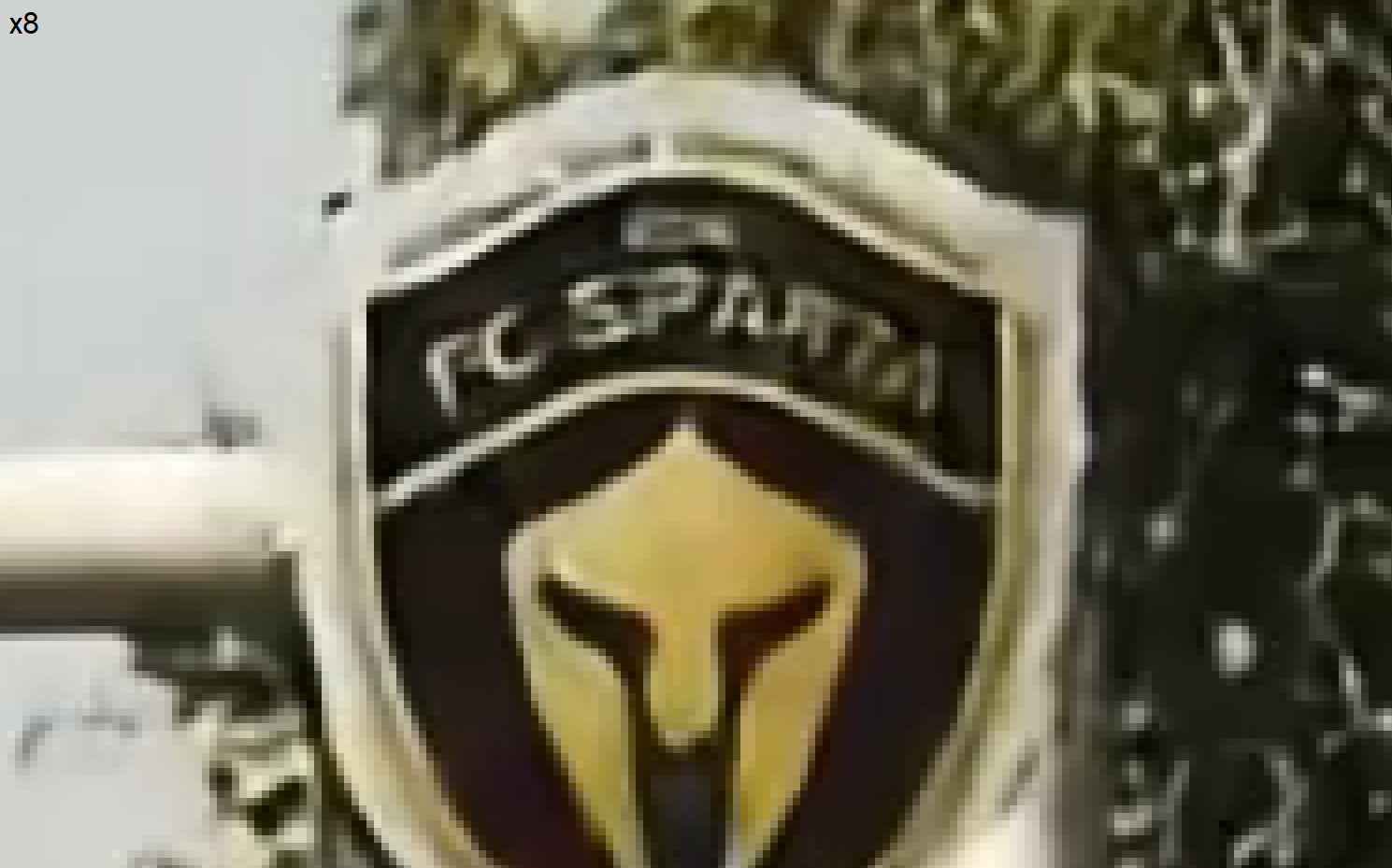}}
    HiNeRV bpp=0.0502 \textbf{(21.43dB)}
    \end{minipage}
    \begin{minipage}[c]{0.24\linewidth}
    \centering
    \centerline{\includegraphics[width=\columnwidth]{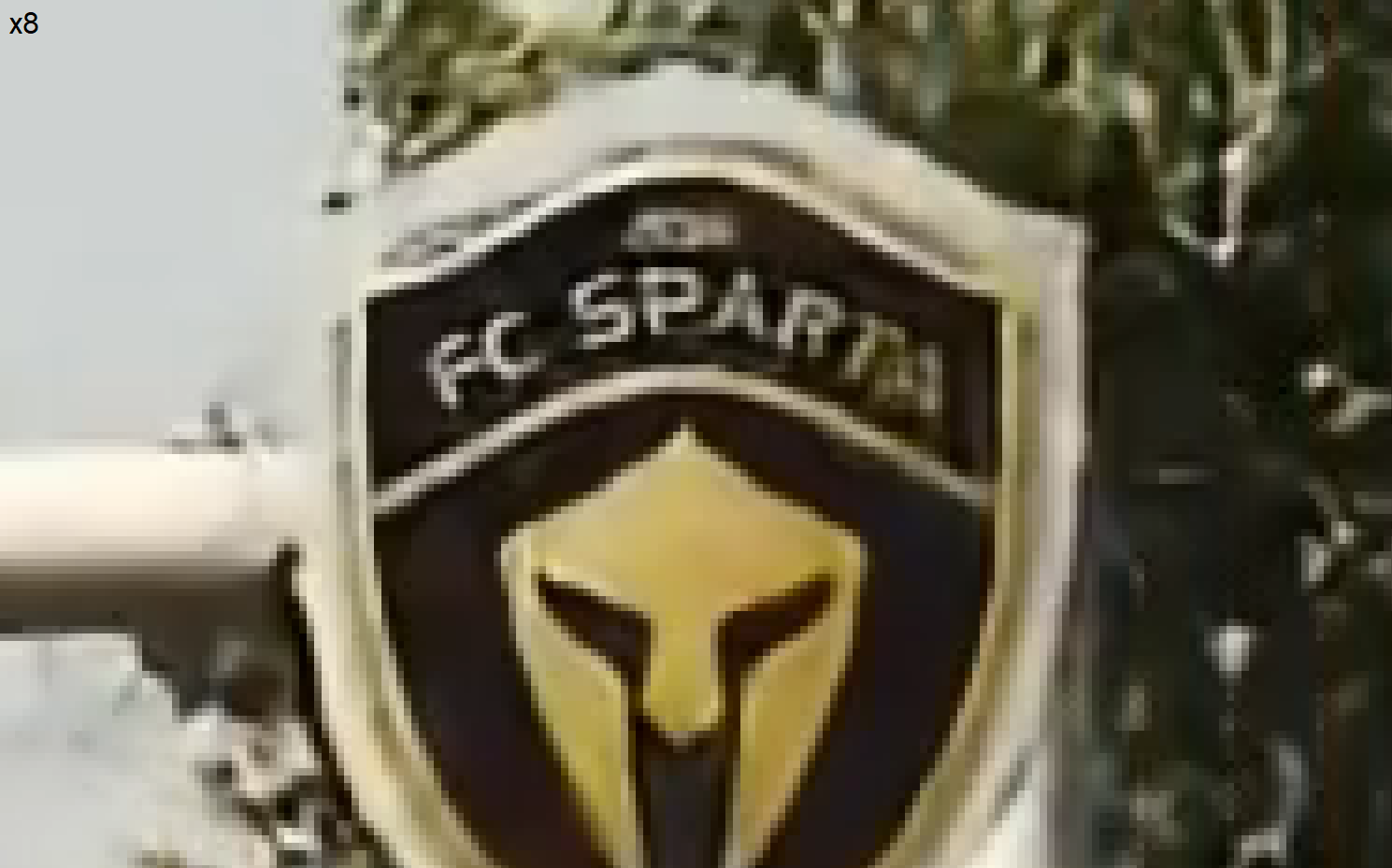}}
    +PT-Loss bpp=0.0501 \textbf{(22.04dB)}
    \end{minipage} 

\caption{Visual comparison between HiNeRV and HiNeRV+PT-Loss. } 
\label{fig_viscompare2}
\vspace{-10pt}
\end{figure*}

To further compare the perceptual quality between the proposed method and the baselines, we have also provided visual comparison examples in \autoref{fig_viscompare1} and \autoref{fig_viscompare2}. In each example, the baseline results and those based on PT-Loss share identical bitrates. It can be observed that for both baseline codecs, the integration of the proposed PT-Loss results in improved perceptual quality, particularly in structured details such as text and edges. These improvements are also verified by higher PSNR/VMAF values measured against the original source sequences.

\subsection{Ablation Study}

To validate the effectiveness of the Mamba-based PT-Loss, we conduct a comparative study against the Swin Transformer-based transcoding quality model (denoted as RankDVQA-UGC) in~\cite{qi2024full}, which has been directly used as a training loss for training DCVC and HiNeRV. \autoref{tab:bdrate_results} reports the BD-rate results on the BVI-UGC dataset across different reference quality groups. For both DCVC and HiNeRV, integrating PT-Loss consistently yields higher overall BD-rate savings compared to using RankDVQA as the perceptual loss, with more evident coding gains in the low-quality reference group. More importantly, PT-Loss is associated with much lower computational complexity compared to RankDVQA-UGC, which makes it more suitable for use as a loss function. Specifically, RankDVQA-UGC, relying on a Swin Transformer-based fusion, yields a complexity of 5.2 GMACs and +4.6M parameters, while our Mamba-based PT-Loss reduces this to 1.4 GMACs and +0.8M parameters.

\begin{figure}[!t]
  \centering
  \begin{minipage}{0.485\linewidth}
      \includegraphics[width=1.1\linewidth]{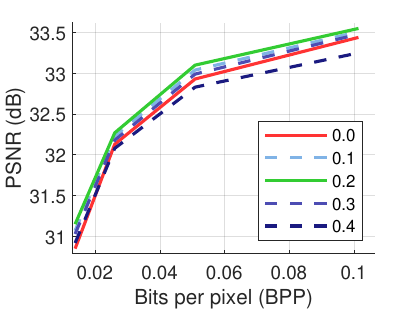}
  \end{minipage}
  \begin{minipage}{0.485\linewidth}
      \includegraphics[width=1.1\linewidth]{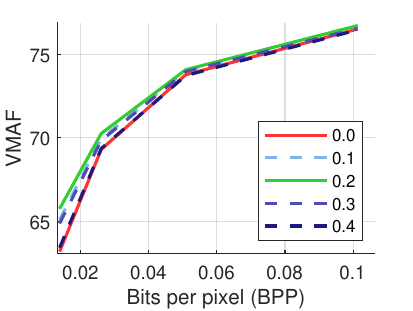}
  \end{minipage}
  \caption{Impact of the perceptual loss weight on overall compression performance when integrated into HiNeRV.}
\label{fig:loss}
\vspace{-15pt}
\end{figure}

To investigate the impact of the perceptual loss weight on compression performance, we conduct an ablation study by varying the weight $\alpha$ in the total loss function. Specifically, we evaluate $\alpha = \{0.0, 0.1, 0.2, 0.3, 0.4\}$ in \autoref{eq:lossHiNeRV} for HiNeRV. Here $\alpha = 0.0$ corresponds to the original HiNeRV setup without PT-Loss. As shown in \autoref{fig:loss}, optimal rate-quality performance for both PSNR and VMAF is achieved when $\alpha = 0.2$. This confirms our selection of $\alpha$.

\section{Conclusion}

In this paper, we have proposed a perceptually-inspired loss specifically targeted at UGC transcoding, which is based on a lightweight, Mamba-accelerated neural network. By redefining the non-pristine reference as a semantic guide rather than employing it as a rigid pixel anchor, our approach enables a host video codec to escape the fidelity trap. Extensive experiments on both autoencoder-based (DCVC) and overfitting-based (HiNeRV) architectures demonstrate that our loss achieves substantial coding gains (up to 12.89\%), particularly when encoding heavily degraded content. Future work should focus on exploring generative compression architectures explicitly designed to reconstruct video quality that surpasses that of the distorted reference.

\small

\bibliographystyle{IEEEtran}
\bibliography{refs}

\end{document}